\documentclass{PoS}

\newcommand{\as}{\alpha_s}

\title{Resummations in PDF fits}

\ShortTitle{Resummations in PDF fits}

\author{\speaker{Marco Bonvini}\\ 
        Rudolf Peierls Centre for Theoretical Physics, 1 Keble Road, University of Oxford, OX1 3NP Oxford, UK\\
        E-mail: \email{marco.bonvini@physics.ox.ac.uk}}

\abstract{In this contribution I discuss recent and ongoing progress on the inclusion of resummation effects in the fit of parton distributions.}

\FullConference{XXIV International Workshop on Deep-Inelastic Scattering and Related Subjects\\
		11-15 April, 2016\\
		DESY Hamburg, Germany}

\begin{document}

Current state-of-the-art PDF sets~\cite
{Ball:2014uwa,Harland-Lang:2014zoa,Dulat:2015mca,Alekhin:2016uxn,Abramowicz:2015mha,Accardi:2016qay,Jimenez-Delgado:2014twa}
are fitted from data assuming fixed-order theory at LO, NLO or NNLO.
The DGLAP evolution is consistently included at LL, NLL or NNLL, respectively.
The quest for precision at the LHC often requires theoretical predictions beyond fixed NNLO.
This can be achieved either with higher fixed-order theory (N$^3$LO), or with the systematic inclusion
of resummed contributions to all orders, or both.
For consistency, PDFs should be fitted with equal (or higher) theoretical accuracy.
While N$^3$LO results for massless DIS are available~\cite{Vermaseren:2005qc}, no N$^3$LO computations are available for the other processes entering
global PDF fits, and DGLAP evolution at N$^3$LL is yet under investigation~\cite{Davies:2016jie}.
On the other hand, resummed computation, which are based on universal features of gluon radiation,
exist for a larger classes of process, and can be used in PDF fits.

The inclusion of resummed computations in PDF fits is not only needed for consistency.
It is generally a beneficial addition, as it is useful to better describe some kinematic regions.
Since PDF fits include data over a wide range in energy $Q$ and momentum fraction $x$,
an accurate theoretical description of extreme regions in both $x$ and $Q$ is important to
improve the quality of the fits.
In this respect, the two most relevant resummations are threshold resummation, describing the large-$x$ region,
and high-energy resummation, describing the small-$x$ region.
Threshold resummation is in principle relevant at all energy scales, even though its effect will be larger at small $Q$
due to the larger $\as$.
High-energy resummation, on the other hand, is mostly important at small $Q$, where data can reach significantly low $x$
(and some tension between fixed-order theory and experimental measurements has been reported~\cite{Caola:2009iy,Caola:2010cy,Abramowicz:2015mha}).
Nevertheless, the effect of small-$x$ resummation in PDF fits will propagate through DGLAP evolution at higher energy scales,
where it will play an important role for High-Energy LHC or a future FCC.

Both threshold and high-energy resummations affect partonic coefficient functions.
While DGLAP splitting functions do not contain any (leading power) large-$x$ logarithmic enhancement,
they contain small-$x$ logarithmic enhancement in the singlet sector.
Therefore, high-energy resummation also affect the evolution of PDFs at small $x$.

For this reason, the inclusion of threshold resummation in PDF fits is somewhat easier,
and has been performed by different groups in the past~\cite{Corcella:2005us,Accardi:2014qda},
even though a global analysis including other processes beyond DIS has been done only recently~\cite{Bonvini:2015ira}.
The latter study, based on the NNPDF~\cite{Ball:2014uwa} fitting methodology, demonstrates the significant effect of threshold resummation
on the PDFs, when comparing a baseline NLO fit with a resummed NLO+NLL fit. However, this effect becomes less significant
when comparing NNLO with NNLO+NNLL. This is partly due to the fact that the current analysis did not include some of the
data usually included in global fits, most importantly inclusive jet cross sections, so that the uncertainty on some PDFs
(mostly the gluon PDF) is larger than in standard global fits.
The inclusion of jet data in resummed PDF fits would reduce the PDF uncertainty, and would highlight
the effect of threshold resummation at NNLO+NNLL.
Therefore, the resummation of jet cross section in PDF fits is a planned future task,
and it will provide a competitive PDF set to be used in LHC analyses for which the description of large $x$ is important
(such as in high-mass SUSY searches~\cite{Beenakker:2015rna}).

Another future development of PDFs with threshold resummation will consist in the use
of improved threshold resummation~\cite{Bonvini:2014joa,Bonvini:2016frm}.
This improved version modifies the form of the large-$x$ logarithms which are resummed,
with the goal of including and mimicking part of the subleading-power large-$x$ logarithms, which are relevant
in intermediate regions at not-too-large $x$. Therefore, the improved threshold resummation effectively predicts part
of the higher order corrections in a region where they behave perturbatively, and can be viewed as a way to approximately
include higher orders.
While the perturbative expansion of a process like DIS has a good convergence pattern (except at large and small $x$),
other processes like top-pair production and jet production are affected by large perturbative corrections:
for these processes, the use of improved threshold resummation will be particularly beneficial.

As far as high-energy resummation is concerned, only one example of PDF fit which includes it exists~\cite{White:2006yh}.
This fit was performed in the DIS factorization scheme, and is based on the Thorne-White (TW) approach
to small-$x$ resummation~\cite{Thorne:1999sg,Thorne:1999rb,Thorne:2001nr,White:2006yh}.
In the 90's and early 2000's, other two groups developed different (though similar) approaches to this resummation:
the Ciafaloni-Colferai-Salam-Stast\`o (CCSS)~\cite{Salam:1998tj,Ciafaloni:1999yw,Ciafaloni:2003rd,Ciafaloni:2007gf}
and the Altarelli-Ball-Forte (ABF)~\cite{Ball:1995vc,Ball:1997vf,Altarelli:2001ji,Altarelli:2003hk,Altarelli:2005ni,Altarelli:2008aj,Rojo:2009us} groups.
Very recently~\cite{Bonvini:2016wki}, the ABF approach was revived (and slightly improved) with the goal of producing a public tool,
named \texttt{HELL} (High-Energy Large Logarithms)~\cite{hell}, which implements the resummation and delivers resummed
splitting functions and coefficient functions.
In particular, the resummation for coefficient functions has been rederived, which leads to a more powerful implementation,
which in turn makes the resummation of other processes much easier than before.
Hence, this tool will be useful to perform resummed global PDF fits.
It has already been interfaced to the \texttt{APFEL} code~\cite{Bertone:2013vaa}, allowing for its use in the NNPDF fitting framework.
Preliminary studies are ongoing, and the production of small-$x$ resummed PDFs within the NNPDF methodology is expected
in the coming months.

\acknowledgments{This work was supported by an European Research Council Starting Grant ``PDF4BSM: Parton Distributions in the Higgs Boson Era''.}

\bibliographystyle{jhep}
\bibliography{references}

\end{document}